\documentclass[aps,prl,twocolumn,groupedaddress]{revtex4}

\usepackage{graphicx}
\begin{document}

\title{Generic differential equation for fractional flow of steady two-phase flow in porous media}


\author{Henning Arendt Knudsen}
\email[]{arendt@phys.ntnu.no}
\noaffiliation{}
\author{Alex Hansen}
\email[]{Alex.Hansen@phys.ntnu.no}
\noaffiliation{}


\date{\today}

\begin{abstract}
We report on generic relations between fractional flow and pressure in steady two-phase flow in porous media. The main result is a differential equation for fractional flow as a function of phase saturation. We infer this result from two underlying observations of steady flow simulations in two and three dimensions using biperiodic boundary conditions. The resulting equation is solved generally, and the result is tested against simulations and experimental relative permeability results found in the literature.
\end{abstract}


\maketitle


Two-phase flow in porous media is possible to classify as either steady or unsteady flow. While steady flow is the one that is statistically invariant in time, the typical unsteady flow is the displacement of one phase by another. In particular, many authors have studied the motion of the displacement front. This was done by experiment\cite{LTZ88}, network modeling\cite{LTZ88}, lattice Boltzmann methods\cite{R90} and also statistical models: diffusion limited aggregation\cite{WS81} and invasion percolation\cite{WW83}. The study of steady flow is often associated with finding the relative permeability curves, although this concept can be defined and is used both for steady and unsteady flow\cite{D92,TD96}. Comparisons between relative permeability obtained by the two methods exist in the literature\cite{PK87}. However, in the complexity of two-phase flow in porous media, pure displacements and true steady flow can be considered to be two limiting situations. Thus, by nature the two situations are very different, and one should expect differences in the relative permeability curves. Nevertheless, unsteady methods are generally preferred because steady methods take long time and are more expensive. The results that we present here concern steady-state flow. A lattice-Boltzmann approach to steady-state flow is found in\cite{LP01}. Further, a very interesting study using a network approach is found in\cite{VP01}.

The body of knowledge in the field is immense\cite{D92,S95}. Pore-network modeling is popular and a good overview of the state-of-art is given in\cite{B01}. Basic mechanisms of displacement on pore level are largely known, and much qualitative knowledge exists about properties on larger scale. However, there is a good deal of work left to do in order to bridge the gap between these two worlds. Numerical simulations is a unique tool to that respect being very well controllable and more precise. In particular, we find that while experimental data points often are few and scattered, we have the possibility to generate sufficient data points to obtain good statistics. Further, in simulation series we can control the \emph{ensemble} completely. Commonly used ensembles for relative permeability or fractional flow curves as functions of saturation, are constant global pressure, constant total flux or constant capillary number(Ca). We argue that the choice of ensemble can be of major importance and is often underestimated. Our results are presented in the formalism of fractional flow and globally applied pressure. We find that when using the constant Ca ensemble, there exist a generic differential equation for the fractional flow. In turn a general solution is found.

The results of the paper is based on a network simulator for immiscible two-phase flow. This line of modeling which is based on Washburn's equation\cite{W21} dates back to the work of several groups in the mid 1980's\cite{KL85,CP91,CP96}. Our model is a continuation of the model developed by Aker \emph{et al.}\cite{AMHB98,AMH98}. Although a thorough presentation can be found in \cite{KAH02}, we provide a brief r{\'e}sum{\'e} of main aspects for clarity.

The porous media are represented by networks of tubes, forming square lattices in two dimensions (2D) and cubic lattices in 3D, both tilted with respect to the imposed pressure gradient and thus the overall direction of flow. We refer to the the lattice points were four(2D) and six(3D) tubes meet as nodes. Volume in the model is contained in the tubes and not in the nodes, although effective node volumes are used in the modeling of transport through the nodes. For further details, see\cite{KAH02}. Randomness is incorporated by distorting the nodes on random within a circle(2D) and a sphere(3D) around their respective lattice positions. This gives a distribution of tube lengths in the system. Further, the radii are drawn from a flat distribution so that the radius of a given tube is $r\in(0.1l,0.4l)\ $where $l\ $ is the length of that tube.

The model is filled with two phases that flow within the system of tubes. The flow in each tube obeys the Washburn\cite{W21} equation, $q=-(\sigma k/\mu)(\Delta p-\sum{p_c})/l$. With respect to momentum transfer these tubes are cylindrical with cross-sectional area $\sigma\ $ and length $l$. The permeability is $k=r^2/8\ $which is known for Hagen-Poiseulle flow. Further, $\mu\ $ is the viscosity of the phase present in the tube. If both phases are present the volume average of their viscosities is used. The volumetric flow rate is denoted by $q\ $and the pressure difference between the ends of the tube by $\Delta p$. The summation is the sum over all capillary pressures $p_c\ $ within the tube. With respect to capillary pressure the tubes are hour-glass shaped meaning that a meniscus at position $x\in(0,l)\ $in the tube has capillary pressure: $p_c=(2\gamma/r)[1-\cos{(2\pi x/l)}]$, where $\gamma\ $is the interfacial tension between the two phases. This is a modified version of the Young-Laplace law\cite{D92,AMHB98}.

Biperiodic(2D) and triperiodic(3D) boundary conditions are used so that the flow is by construction steady flow. That is to say, the systems are closed so both phases retain their initial volume fractions, i.e. their saturations. These boundary conditions are in 2D equivalent to saying that the flow is restrained to be on the surface of a torus. The flow is driven by a globally applied pressure gradient. Usually invasion processes are driven by setting up a pressure fall between two borders, inlet and outlet. Since, by construction, the outlet is directly joined with the inlet in our system, we give instead a so-called global pressure drop when passing this line or cut through the system\cite{roux,KAH02}. Effectively we put restraints on the pressure gradient that is experienced throughout the network. Integration of the pressure gradient along an arbitrary closed path one lap around the system, making sure to pass the `inlet-outlet'-cut once, should add up to the same global pressure fall.

For a given value of the global pressure fall, and given the instantaneous distribution of the phases in the system, the pressure field is calculated numerically, and the flow distribution at that moment is known. Given the flow field the whole system is updated according to the Euler scheme. All interfaces, menisci, which are the essential entities keeping track of the distribution of the phases, are moved according to the fluid velocity in the tube where they are situated. After having updated the hole system, the new distribution of the phases might give new effective viscosities in the tubes and different capillary pressures across the menisci, leading to a recalculation of the coefficients in the set of equations for the pressure. The tricky part is how to model the transport of the menisci across the nodes, and this is done by applying a set of rules that assures volume conservation of both phases and that are acceptably close to the real world. We refer to the previous study for details on this\cite{KAH02}.

All simulations being the basis for our conclusions are done in series where the saturation is varied as the independent variable and the capillary number as well as the viscosity ratio $M=\mu_{\rm nw}/\mu_{\rm w}\ $is held fixed. We define the capillary number as ${\rm Ca}=Q_{\rm tot}\mu_{\rm eff}/(\Sigma\gamma)\ $ where $\Sigma\ $ is the cross-sectional area of the entire network, $Q_{\rm tot}\ $is the total flux through the network and $\mu_{\rm eff}=\mu_{\rm nw}F_{\rm nw}+\mu_{\rm w}F_{\rm w}$. Here, $F_{\rm nw}\ $ and $F_{\rm w}\ $are nonwetting and wetting fractional flow respectively. In invasion studies one often employs either the wetting or the nonwetting viscosity and refer to the wetting or nonwetting capillary number, respectively. However, in steady flow it is appropriate to take the volume average. The resulting dependent variables are the fractional flow of the two phases and the globally applied pressure. These two variables are in a one-to-one relationship with the two relative permeabilities, but we will stick to the former notation here, for clarity and convenience.

From the simulations two observations are made from which we infer the following differential equation for the fractional flow as a function of saturation;
\begin{equation}
\label{eq:diff}
\frac{d^2F}{dS^2}-(2aF+b)\frac{dF}{dS}=0.
\end{equation}
The saturation is the nonwetting saturation $S=S_{\rm nw}\ $and the fractional flow is the nonwetting fractional flow $F=F_{\rm nw}(S_{\rm nw})\ $here and in illustrations unless where explicitly stated otherwise. However, the relation is equivalent in form for the wetting counterpart $F_{\rm w}(S_{\rm w})$, only the values of $a\ $and $b\ $change.

\begin{figure}
\includegraphics[width=8cm]{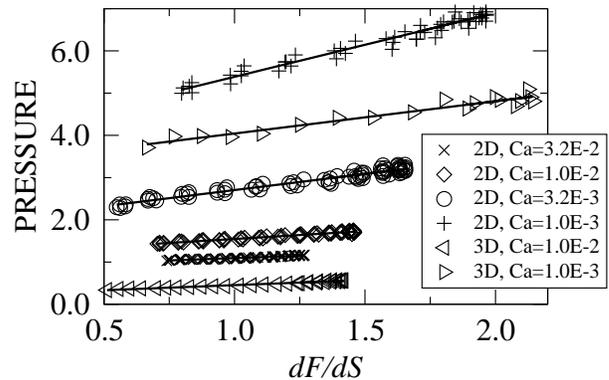}
\caption{\label{fig:dfdsfit}
The figure shows that the relation between pressure and the derivative of the fractional is of the form in Eq. (\ref{eq:dfds}). The curves are for viscosity matching phases for four different capillary numbers in 2D and two capillary numbers in 3D. The nonwetting saturation $S_{\rm nw}\ $serves as parameter for the curves, and the intervals in which the data is plotted are as follows; (2D) ${\rm Ca}=3.2\times10^{-2}:S_{\rm nw}\in(0.14,0.93)$,${\rm Ca}=1.0\times10^{-2}:S_{\rm nw}\in(0.14,0.83)$,${\rm Ca}=3.2\times10^{-3}:S_{\rm nw}\in(0.14,0.83)$,${\rm Ca}=1.0\times10^{-3}:S_{\rm nw}\in(0.20,0.73)$, (3D) ${\rm Ca}=1.0\times10^{-2}:S_{\rm nw}\in(0.10,0.90)$,${\rm Ca}=1.0\times10^{-3}:S_{\rm nw}\in(0.20,0.75)$. The two curves in 3D are lowered by one unit for clarity.
}
\end{figure}

The two observations on which Eq. (\ref{eq:diff}) is based are of a phenomenological character. The physical problem is intractable from an analytical point of view. By means of simulations the evolution and properties are monitored in detail. The first observation is that the derivative of the fractional flow is related to global pressure by
\begin{equation}
\label{eq:dfds}
P(S)=A\frac{dF(S)}{dS}+B,
\end{equation}
where $P\ $ denotes normalized pressure, i.e., the actual pressure for the fixed flux at a given saturation divided by the pressure for single phase flow at the same flux. This result has been reported on in great detail for the case of viscosity matching phases\cite{KH02}. Those results were solely based on simulations in two dimensions. However, the result is extendable to 3D, see Fig.\ref{fig:dfdsfit}, as well as viscosity contrasts.

\begin{figure}
\begin{tabular}{ll}
\includegraphics[width=4cm]{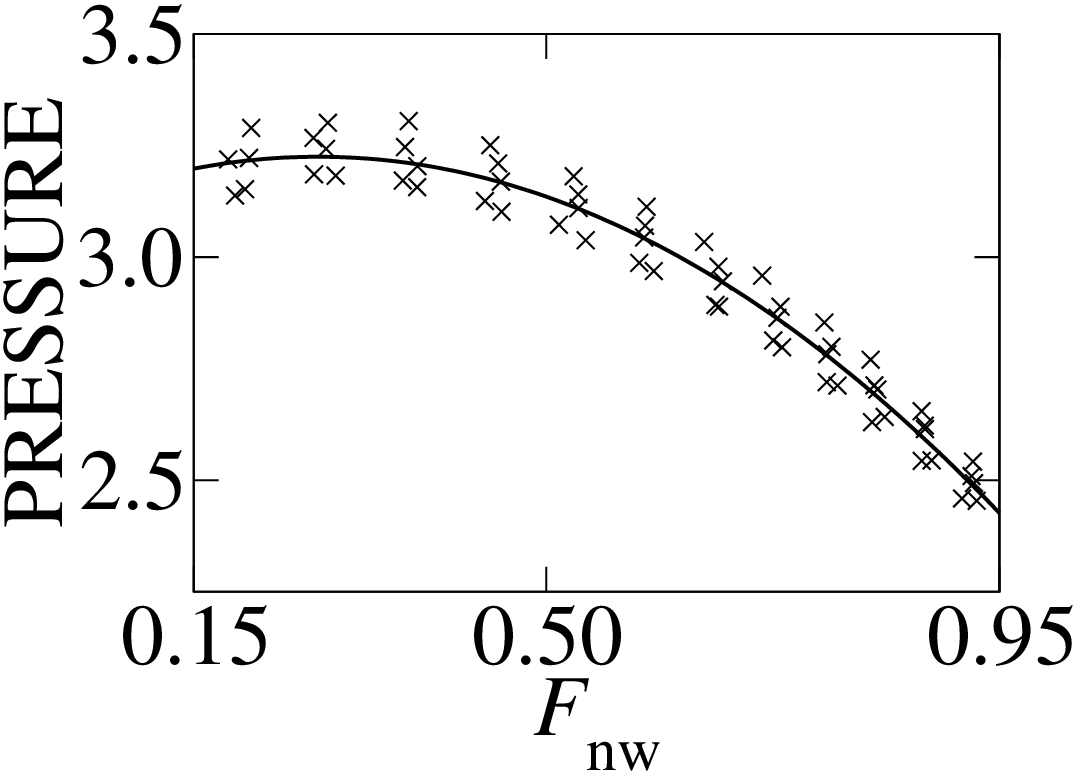}
&
\includegraphics[width=4cm]{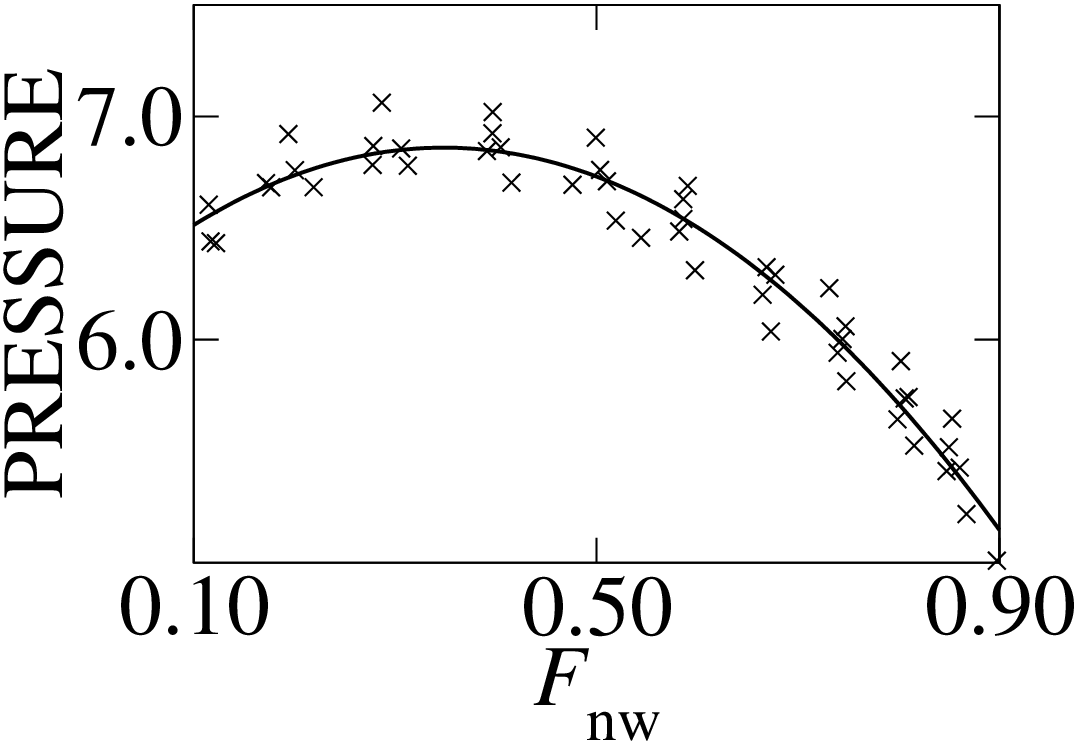}
\\
(a) ${\rm Ca}=3.2\times10^{-3}$ &(b) $ {\rm Ca}=1.0\times10^{-3}$
\\
\includegraphics[width=4cm]{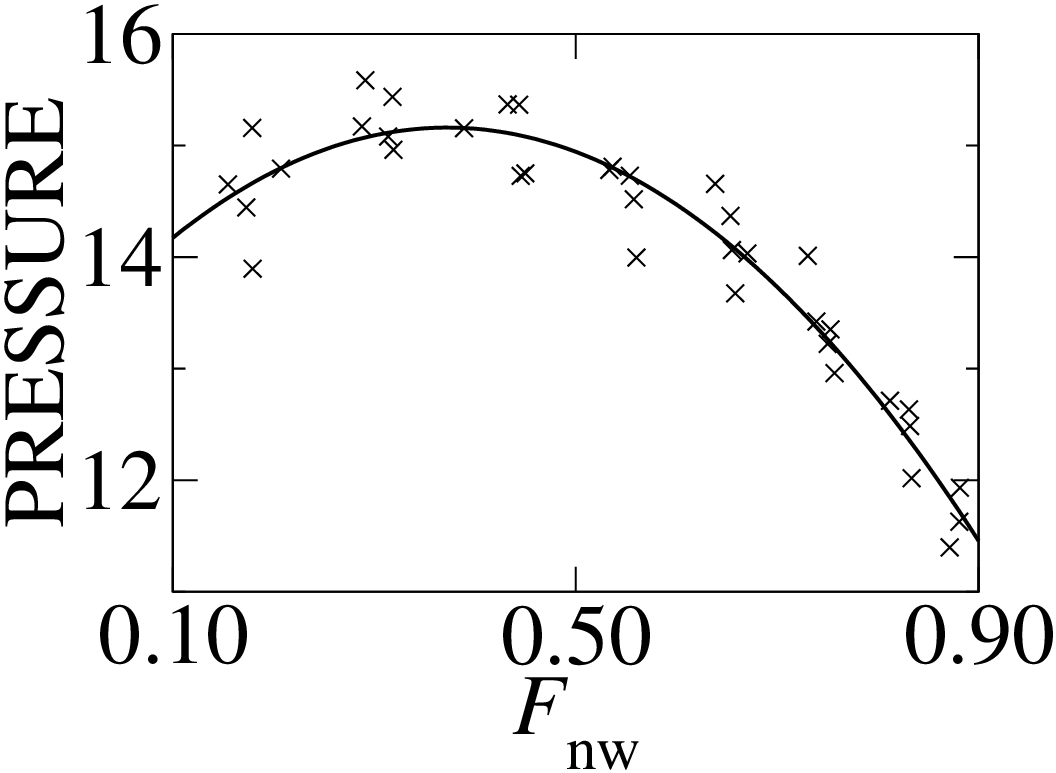}
&
\includegraphics[width=4cm]{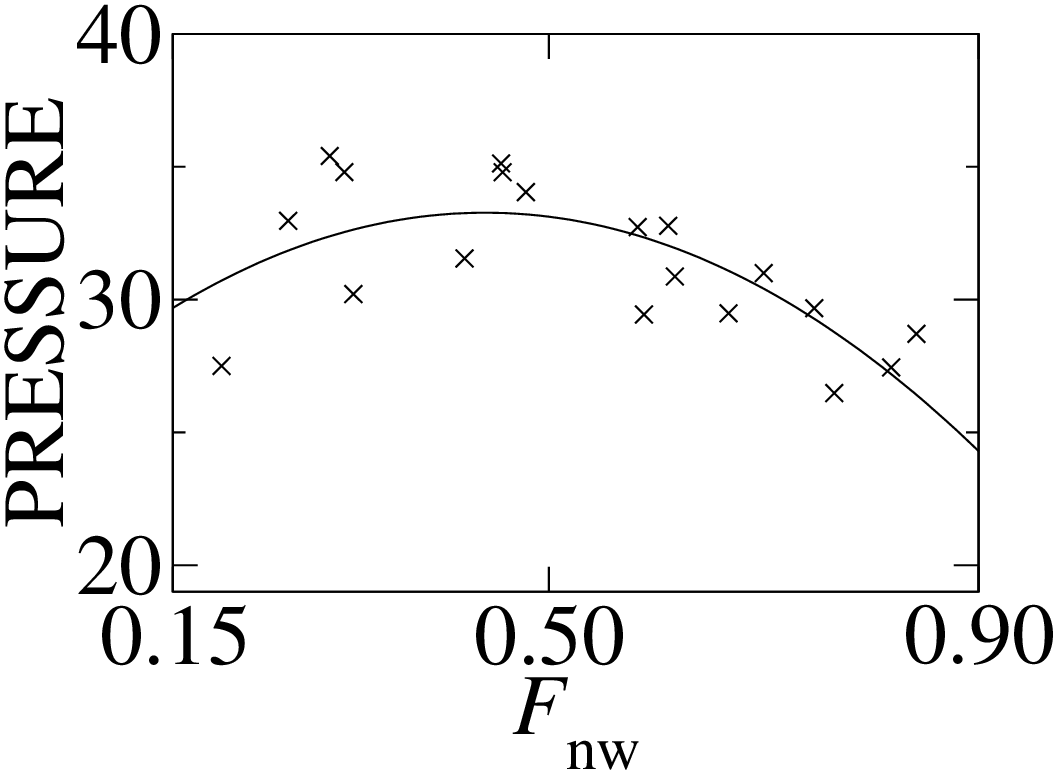}
\\
(c) ${\rm Ca}=3.2\times10^{-4}$ &(d) $ {\rm Ca}=1.0\times10^{-4}$
\end{tabular}
\caption{\label{fig:pffit}
The figure shows fitted curves to $P(F_{\rm nw})\ $according to Eq. (\ref{eq:qvadr}). The results are from the same series of simulations as the curves in 2D in Fig. \ref{fig:dfdsfit}, only showing results for two lower capillary numbers. The shown region of $F_{\rm nw}\ $is the one where the fit is good. For these capillary number this region is large, for the two higher values of ${\rm Ca}\ $ shown if Fig. \ref{fig:dfdsfit}, the region of good fit is; ${\rm Ca}=3.2\times10^{-2}:F_{\rm nw}\in(0.15,0.45)$,${\rm Ca}=1.0\times10^{-2}:F_{\rm nw}\in(0.10,0.50)$. The data is from five realizations of the porous network giving slightly shifted curves as we can see. The scattering of the data becomes larger for smaller capillary numbers due to increased hysteresis and history effects.
}
\end{figure}

\begin{figure}
\includegraphics[width=8cm]{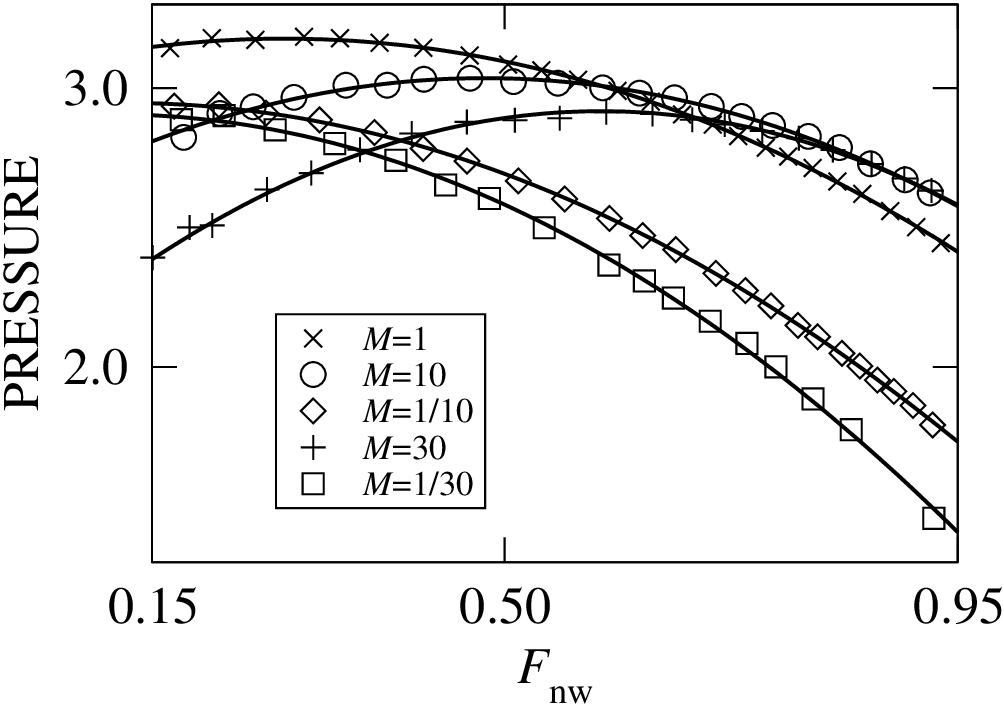}
\caption{\label{fig:pfLfit}
We show $P(F_{\rm nw})\ $for a fixed value of the capillary number, namely ${\rm Ca}=3.2\times10^{-3}$, which corresponds to Fig. \ref{fig:pffit}(a). We study one sample assigning five different viscosity ratios $M\ $to the phases. We observe how it is possible to fit all the curves to the form in Eq. (\ref{eq:qvadr}) for $M\in(1/30,30)$.
}
\end{figure}

The second observation is that for a broad range of saturation values the pressure forms a quadratic function of the fractional flow. In a mathematical formulation this result becomes
\begin{equation}
\label{eq:qvadr}
P(F)=a'F^2+b'F+c'.
\end{equation}
The range of validity of this equation is illustrated by the quadratic fits in Figs. \ref{fig:pffit} and \ref{fig:pfLfit}. In Fig. \ref{fig:pffit} the viscosity ratio is unity, and we observe how Eq. (\ref{eq:qvadr}) holds for each fixed value of ${\rm Ca}$. Likewise, we show in Fig. \ref{fig:pfLfit} quadratic fits to $P(F)$-curves for one given capillary number ${\rm Ca}=3.2\times10^{-3}$, varying the viscosity ratio from $M=30\ $to $M=1/30$. The point is that for most of the range that we are interested in, it is sufficient to expand $P(F)\ $to second order in $F\ $around the maximum point. Alternatively, an expansion of $P(S)\ $in $S\ $ turns out to require higher order terms, and is hence less useful. Other combinations of capillary number and viscosity ratio have been performed and they are found to agree with this result. The shifting of the curves and detailed dependence on ${\rm Ca}\ $and $M\ $ are to be discussed elsewhere\cite{KHunpub02}.

\begin{figure}
\includegraphics[width=8cm]{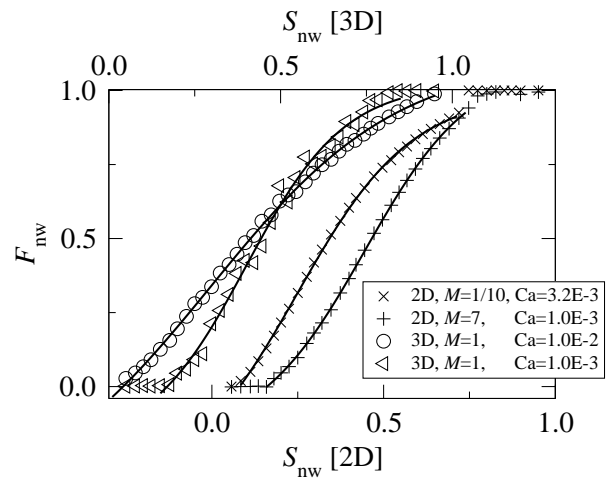}
\caption{\label{fig:fsLg3fit}
The figure shows how four samples of fractional flow curves from our simulations can be fitted to the general solution of Eq. (\ref{eq:diff}) as it is given in Eq. (\ref{eq:solution}). The dimensionality, viscosity ratio and capillary number of these samples are as indicated in the legend. For clarity the curves are shifted so that the two curves in 2D belong the lower x-axis, and the two curves in 3D belong to the upper x-axis. The 3D curves are from the same simulation series as the 3D curves in Fig.\ref{fig:dfdsfit}.
}
\end{figure}

\begin{figure}[!h]
\includegraphics[width=8cm]{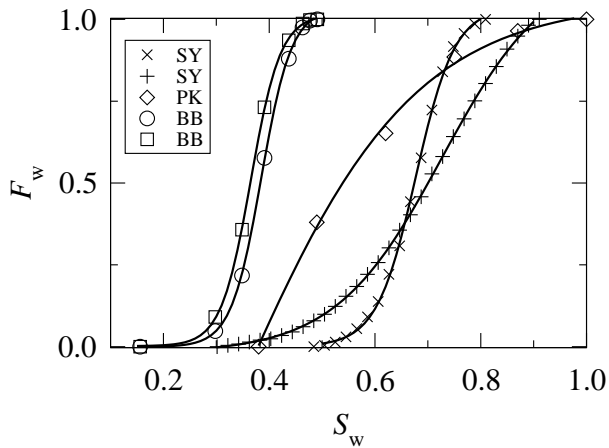}
\caption{\label{fig:PKandSYfit}
Data from the literature: Sharma and Yen(SY)\cite{SY83}, Peters and Khataniar(PK)\cite{PK87},Braun and Blackwell(BB)\cite{BB81,D92}. The water saturation is also the wetting saturation $S_{\rm w}$, but note that it is partly the different normalization of this entity, that was employed in the respective papers, which decide the placement of the curves. The SY-curves are typical samples of the set of functional forms that are employed when discussing how various physical parameters influence on fractional flow. Regarding PK and BB-curves, see Fig.\ref{fig:PKdfP}. All the curves are fitted to the solution of Eq.(\ref{eq:diff}) as it is given in Eq.(\ref{eq:solution}).
}
\end{figure}

\begin{figure}[!h]
\includegraphics[width=8cm]{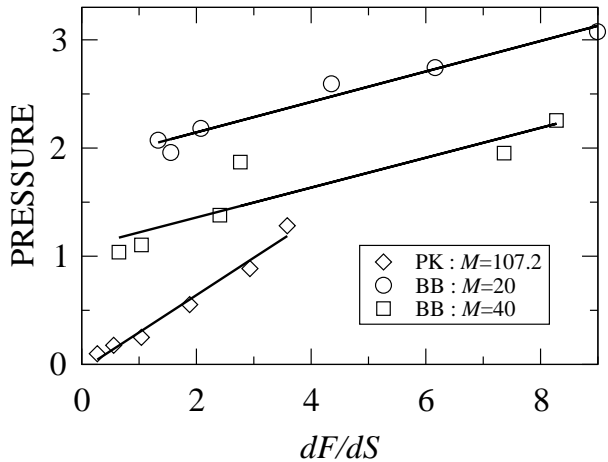}
\caption{\label{fig:PKdfP}
The curves correspond to the respective curves of Fig.\ref{fig:PKandSYfit}. For PK the reported value of the viscosity ratio $M=107.2\ $is used for the calculation from relative permeabilities to fractional flow and pressure. However, for BB we have chosen two different reasonable values for the viscosity ratio, to illuminate its effect. 
}
\end{figure}

Recall that $S\ $ is the independent variable upon which both $F\ $ and $P\ $depend. The two Eqs.(\ref{eq:dfds}) and (\ref{eq:qvadr}) for pressure $P\ $must be equal for all saturations $S$. Combining the two gives a first order differential equation that differentiated once again with respect to $S\ $ becomes Eq.(\ref{eq:diff}). Note that in Eq. (\ref{eq:diff}) $a=a'/A\ $ and $b=b'/A$. The second order version being more elegant is, however, autonomous and just as simple to solve using standard methods\cite{BO78}. The general solution is
\begin{equation}
\label{eq:solution}
F(S)=-\frac{1}{2a} \frac{(b-k)+(b+k)\exp{[k(S-S_0)]}}{1+\exp{[k(S-S_0)]}},
\end{equation}
where
\begin{equation}
k^2=b^2-4ak_0 ,
\end{equation}
and $k_0\ $and $S_0\ $are constants of integration. This result is generic and general. In Fig.\ref{fig:fsLg3fit} we provide the fractional flow by simulations for four different sets of parameters. The sets are fitted by the form in Eq.(\ref{eq:solution}). We observe that the fits are excellent. Even though the two underlying observations are valid only in a certain central region (say 70-90\%) of the curves, the solution is satisfactory in the entire two-phase region.

Relative permeability is used both for steady-state flow and unsteady flow. The literature covers several methods to produce relative permeability from experimental data\cite{D92,TD96}. It is a priori not entirely clear to which extent a given study is comparable to our simulations. We have chosen a few samples from literature which are provided in Figs.\ref{fig:PKandSYfit} and \ref{fig:PKdfP}. In literature, when focus is on how some physical or chemical properties influence on fractional flow, this is often done by employing a class of functional forms for the fractional flow. The curves indicated by SY are samples of that\cite{SY83}. It is reassuring that these curves can be fitted by Eq.\ref{eq:solution}. The other three curves are calculated from the relative permeabilities $k_{\rm rnw}\ $and $k_{\rm rw}\ $and the viscosity ratio $M\ $using the formalism $P = (k_{\rm rnw}+Mk_{\rm rw})^{-1}\ $and $F_{\rm nw} = Mk_{\rm rw} P$, which arise when assuming that the total flux is held constant under the series and $P\ $is normalized pressure, for details see\cite{D92,KH02}.

The PK-curve is a steady-state curve which is comparable to our simulations, even though the ensemble is constant total flux. The validity of the Eq.(\ref{eq:solution}) follows from Fig.\ref{fig:PKandSYfit}, likewise Eq.(\ref{eq:dfds}) follows from Fig.\ref{fig:PKdfP}. Here we use the actual viscosity ratio that was used in the experiments for our calculations. It is well established that relative permeability is a function of a large number of parameters including the viscosity ratio\cite{VP01}. However, to some extent one gets the impression that the curves are regarded as rock properties and are valid for at least a range of viscosity ratios. This might be very system dependent. The curves marked BB\cite{BB81,D92} in Figs.\ref{fig:PKandSYfit} and \ref{fig:PKdfP} are generated from relative permeability data in this way by choosing two values for viscosity ratio: $M=20\ $and $M=40$. Within this range the results are robust.

In general fractional flow curves are obtained for constant total flux. Another possible ensemble is constant pressure drop over a sample. In our simulations we have chosen a third ensemble, namely constant capillary number. By this method we keep the ratio between viscous and capillary forces fixed. The precision and control of parameters we obtain in the curves exceed what is normal in experiments. This is highly advantageous when looking for general relationships between variables. We believe that constant Ca is the appropriate ensemble for the study of steady flow properties. The comparisons with literature in this study indicate the correctness of the result, however, carefully designed laboratory experiments should be made using this ensemble to check our results.

In conclusion it is our claim that this general result is valid for all steady-state two-phase flow system that can be modeled by our numerical model. That is to say immiscible flow where film flow can be neglected. This is more likely to be case at moderate to high capillary numbers. Studies of imbibition and drainage by small scale experiments show that in particular imbibition like steps may be dominated by film flow at low capillary numbers, but not at high capillary numbers\cite{CK85}.

\begin{acknowledgments}
H.A.K. acknowledges support from VISTA, a collaboration between Statoil and The Norwegian Acad. of Science and Letters. This work has received support from NTNU through a grant of computing time on the good supercomputing facilities at NTNU.
\end{acknowledgments}

\bibliography{referanser}

\end{document}